\documentclass[final,3p,times,authoryear]{elsarticle}

\usepackage{amssymb}
\usepackage{amsmath}

\usepackage[T1]{fontenc}
\usepackage{url}

\journal{Planetary and Space Science}

\begin{document}

\begin{frontmatter}



\title{On ore-bearing asteroid remnants in lunar craters}

\author[indres]{Jayanth Chennamangalam\corref{cor1}} 
\author[birm]{Paul Brook} 
\author[hscfa]{Martin Elvis}
\author[erau]{Samuel Peterson}

\cortext[cor1]{jayanth@jayanthchennamangalam.com}

\affiliation[indres]{organization={Independent researcher},
            city={Vancouver},
            state={BC},
            country={Canada}
}
\affiliation[birm]{organization={Institute for Gravitational Wave Astronomy and School of Physics and Astronomy, University of Birmingham},
            addressline={Edgbaston},
            city={Birmingham},
            postcode={B15 2TT},
            country={UK}}
\affiliation[hscfa]{organization={Center for Astrophysics | Harvard \& Smithsonian},
            addressline={60 Garden Street},
            city={Cambridge},
            postcode={MA 02138},
            country={United States}}
\affiliation[erau]{organization={Embry-Riddle Aeronautical University},
            addressline={1 Aerospace Boulevard},
            city={Daytona Beach},
            postcode={FL 32114},
            country={United States}}

\begin{abstract}
We modify the probabilistic formalism developed by \citet{elvis2014} to estimate the number of lunar craters that contain ore-bearing asteroid remnants. When we consider craters at or above a threshold diameter of 1\,km, we estimate an upper limit of ${\sim}6,500$ craters with asteroid remnants containing significant amounts of platinum group metals and an upper limit of ${\sim}3,400$ craters with asteroid remnants that contain significant amounts of water in the form of hydrated minerals. For a more conservative threshold of 5\,km, we estimate $\lesssim400$ craters with asteroid remnants that contain significant amounts of platinum group metals. These values are one to two orders of magnitude larger than the number of ore-bearing near-Earth asteroids estimated by \citet{elvis2014}, implying that it may be more advantageous, and hence more profitable, to mine asteroids that have impacted the Moon rather than the ones that are in orbit.
\end{abstract}




\end{frontmatter}



\section{Introduction}
\label{sec:intro}

Extracting resources from Solar System bodies has been a topic of scientific interest for several decades \citep[see, e.g.,][]{lewis1996}. The past two decades witnessed the birth, and subsequent demise, of the first commercial enterprises with the explicit aim of mining asteroids\footnote{Planetary Resources and Deep Space Industries, for example.}. More recently, in tandem with the increasing commercialization of space, there has been a renewed interest in the topic, with companies focusing on mining platinum group metals (PGMs) and water from near-Earth asteroids (NEAs) and helium-3 and water from the Moon.

Based on optical spectra, radar albedos, and studies of iron meteorites thought to derive from them, M-type asteroids are believed to be metal-rich, made up mostly of iron and nickel, with a variety of siderophile elements dissolved in them, including PGMs. Similarly, C-type asteroids are believed to contain hydrated minerals. For the purposes of mining, the material that makes up these asteroids are considered to be `ores' if it is commercially viable to extract PGMs or water from them. To address the question of how many ore-bearing NEAs exist that we can realistically access and profitably mine, \citet{elvis2014} has developed a probabilistic formalism (see Section~\ref{sec:estimation}) which yields values of ${\sim}10$ for PGM ore-bearing asteroids and ${\sim}18$ for water-bearing ones.

In this paper, we consider the case of asteroids that have struck the Moon. It is well-known that the lunar regolith contains exogenously delivered material, originating from the solar wind along with asteroids and comets that have collided with the Moon, producing its many craters \citep[see, e.g.,][]{crawford2023}. It is generally expected that while small fragments of the impactor survive the collision, most of it vaporizes and is dispersed over a large area. However, recent work has suggested that for a fraction of collisions, most of the asteroid material survives, albeit deformed \citep{yue2013,svetsov2015,halim2024}. \citet{yue2013} perform numerical simulations to find that for vertical impact velocities $< 12$\,km\,s$^{-1}$, most of the asteroid survives the collision, and, in the case of complex craters, is swept into the center of the crater. In the case of simple craters, which are smaller, the material is mixed into the breccia lens that fills the crater.

A simple estimate for the number of lunar craters containing PGM ore is provided by \citet{wingo2004}. The figure of approximately 28,000 is based on the then available count of craters $\geq 1$\,km in diameter and assumes that  M-type NEAs created 3\% of those craters. In this paper, we apply a modified version of the \citet{elvis2014} formalism to the Moon to obtain a refined estimate of the number of craters with asteroid remnants that contain PGM ore, and an estimate of the number of craters with water-bearing remnants.

The organization of this paper is as follows. In Section~\ref{sec:estimation} we introduce the formalism used to estimate the number of lunar craters containing ore-bearing asteroid remnants. In Section~\ref{sec:pgms}, we use the formalism to estimate the number of PGM ore-bearing craters, and in Section~\ref{sec:water}, we use it to obtain the number of water-bearing craters, before concluding in Section~\ref{sec:discussion}.

\section{Estimating the number of ore-bearing lunar craters}
\label{sec:estimation}

\citet{elvis2014} quantifies the number of asteroids that contain a given resource as
\begin{equation}
N_{\rm ore} = P_{\rm type} \times P_{\rm rich} \times P_{\rm acc} \times P_{\rm eng} \times N(> M_{\rm min}),
\label{eq:elvis}
\end{equation}
where $P_{\rm type}$ is the probability that an asteroid is of the spectral type that allows for the existence of the resource in it, $P_{\rm rich}$ is the probability that the asteroid is sufficiently rich in the resource to make extraction profitable, $P_{\rm acc}$ is the probability that the asteroid is accessible, i.e., the delta-v required to reach the asteroid and return with mined material is achievable, $P_{\rm eng}$ is the probability that the engineering challenges of mining are surmountable, and $N(> M_{\rm min})$ is the number of asteroids that have a mass larger than a threshold $M_{\rm min}$ to be profitable to mine.

We modify Eq.~(\ref{eq:elvis}) in the following way in order to apply it to lunar craters. Since the entire surface of the Moon is accessible to human spacecraft, $P_{\rm acc} = 1$, so this term is removed from the equation. For NEAs this was the dominant term.

To quantify the fact that not all asteroids that impact the Moon survive, a new term, the probability of the asteroid surviving in the impact crater, $P_{\rm surv}$, is introduced and is discussed in the next section.

Combining these terms gives the number of ore-bearing asteroids whose remnants survive in lunar craters,
\begin{equation}
N_{\rm ore} = P_{\rm type} \times P_{\rm rich} \times P_{\rm surv} \times P_{\rm eng} \times N(> M_{\rm min}).
\label{eq:mod_elvis}
\end{equation}

Masses of NEAs are measurable only in certain circumstances, e.g., in the case of binary asteroids. \citet{elvis2014} therefore assumes a mean asteroid density so that the observable quantity $N(> D_{\rm min})$ can be used as a proxy for $N(> M_{\rm min})$, where $D_{\rm min}$ is the minimum diameter for profitability, which depends on the resource under consideration. We also need to use a proxy for the mass threshold $M_{\rm min}$, but the diameter of asteroids that created lunar craters are also not observable in our scenario. The only quantity that we can directly measure is the diameter of the crater resulting from the collision. Therefore, we use $N(> D_{\rm c,min})$ as a proxy for the last term, where $D_{\rm c,min}$ is the minimum crater diameter. In the following analyses, we assume particular values for $D_{\rm c,min}$ for the two cases of PGM ore-bearing and water-bearing asteroids.


\subsection{Platinum group metals}
\label{sec:pgms}

For PGMs, \citet{elvis2014} applies Eq.~(\ref{eq:elvis}) to M-type NEAs, assuming the values $P_{\rm type} = 0.04$, $P_{\rm rich} = 0.5$, $P_{\rm acc} = 0.025$, $P_{\rm eng} = 1$, and $N(> D_{\rm min}) = 2 \times 10^4$. $P_{\rm eng}$ includes all the engineering details related to ore extraction and is difficult to assess, so it is assumed to be unity, rendering $N_{\rm ore}$ an upper limit. \citet{elvis2014} takes $D_{\rm min}$ to be 100\,m to allow for the asteroid to be worth roughly US\$1\,billion at 2014 PGM prices\footnote{US\$50,000/kg. At the time of writing, the price of platinum is US\$33,000/kg, but this does not change the diameter threshold significantly.}.

We adopt the same values of $P_{\rm type}$ and $P_{\rm rich}$ for Eq.~(\ref{eq:mod_elvis}). This is reasonable because lunar craters were formed by collisions with near-Earth objects.

$P_{\rm surv}$ is more complex, being a combination of resource type, impact speed, and impact angle. \citet{yue2013} perform hydrodynamic simulations to find that for vertical impact velocities $< 12$\,km\,s$^{-1}$, ``much'' of the asteroid survives the collision, although they do not report the fraction that does. The distribution of the remnants within the resulting crater depends on whether the crater is simple or complex. Complex craters are large --- typically $\gtrsim19$\,km in diameter on the Moon \citep{kruger2018} --- and contain a central feature such as a peak or a ring. \citet{yue2013} show that in such craters, most of the impactor material is swept into the central feature, although they do not specify the fraction. Simple craters are smaller and bowl-shaped. In this case, the impactor material is mixed into the breccia lens that fills the crater. The amount of impactor material that is found inside the crater depends on the impact angle. \cite{bland2008} simulate impacts with velocities $\leq 7$\,km\,s$^{-1}$ for various impact angles, and report that for angles $45^\circ$ and above, $\geq 42$\% of the asteroid remains in the crater. For more oblique impacts, the material is dispersed down range from the crater \citep[see also][]{halim2024}. \citet{svetsov2015} conduct simulations that show that at the most probable impact angle of $45^\circ$ \citep{shoemaker1961}, depending on the projectile and target materials, 10-80\% of the asteroidal mass remains unmelted at an impact velocity of 12\,km\,s$^{-1}$, and larger fractions remain unmelted at lower velocities. We choose 12\,km\,s$^{-1}$ as the threshold velocity below which a significant fraction of the asteroid survives, and, based on the distribution of vertical impact velocities presented in \citet{yue2013}, we take $P_{\rm surv} = 0.25$. We do not consider the distribution of impact angles in this analysis, but note that $P_{\rm surv}$ will, in practice, be a function of not just the impact velocity but also the angle, in addition to the material properties of both the projectile and the target.

We continue to use $P_{\rm eng} = 1$, noting that this results in $N_{\rm ore}$ being an upper limit.

The relationship between crater sizes and that of the objects that create them is not straightforward \citep[see, e.g.,][]{collins2005,marchi2009,johnson2016}. \citet{collins2005} give the diameter of the transient crater formed on impact as

\begin{equation}
D_{\rm tc} = 1.161 \left(\frac{\rho_{\rm i}}{\rho_{\rm t}}\right)^{1/3} D_{\rm i}^{0.78} v_{\rm i}^{0.44} g^{-0.22} \sin^{1/3}{\theta_{\rm i}},
\end{equation}

where $\rho_{\rm i}$ and $\rho_{\rm t}$ are the densities of the impactor and target, respectively, $D_{\rm i}$ is the diameter of the impactor, $v_{\rm i}$ is the impact velocity, $g$ is the acceleration due to gravity at the surface of the target, and $\theta_{\rm i}$ is the impact angle (all quantities in SI units). Shortly after the formation of the transient crater, its walls collapse to form the final crater. For simple craters, the final diameter is given by \citet{collins2005} as

\begin{equation}
D_{\rm c} \approx 1.25 D_{\rm tc}.
\end{equation}

On the Moon, whose surface gravity is 1.63\,m\,s$^{-2}$, assuming $\rho_{\rm i} = \rho_{\rm t}$, an asteroid of diameter 100\,m impacting at a velocity of 12\,km\,s$^{-1}$ creates a crater of diameter ${\sim}$1--3\,km. The actual value depends on the impact angle. We therefore compute $N_{\rm ore}$ for multiple values of $D_{\rm c,min}$: 1\,km, 3\,km, and 5\,km. The fact that some fraction of the impactor gets vaporized motivates our choice for the largest of these thresholds. There are $N(> D_{\rm c,min}) = 1,296,796$ craters $\geq 1$\,km in diameter\footnote{\citet{robbins2018} reports 1,296,879 craters $\geq 1$\,km in diameter, but the corresponding database lists only 1,296,796. We use the latter value in our calculations. This does not affect the results significantly.}, 212,149 that are $\geq 3$\,km in diameter, and 83,061 craters that are $\geq 5$\,km in diameter \citep{robbins2018}. For $D_{\rm c,min} = 1$\, km, Eq.~(\ref{eq:mod_elvis}) gives
\begin{equation*}
\begin{split}
N_{\rm ore} &= 0.04 \times 0.5 \times 0.25 \times 1 \times 1,296,796 \\
&= 6,484.
\end{split}
\end{equation*}

For crater diameter thresholds of 3\,km and 5\,km, we obtain $N_{\rm ore} = 1,061$ and 415, respectively.

Considering only complex craters in which the asteroid material is swept into the central feature, and taking the simple-to-complex transition diameter of $19$\,km \citep{kruger2018} as $D_{\rm c,min}$, we obtain $N_{\rm ore} = 38$.

These results are summarized in Table~\ref{tab:estimates}. 

\begin{figure}[t]
    \centering\includegraphics[width=0.5\linewidth, alt={Log-log plot with crater diameter threshold on the X-axis and the number of ore-bearing craters on the Y-axis. The points lie roughly along a line with a slope of approximately -2.}]{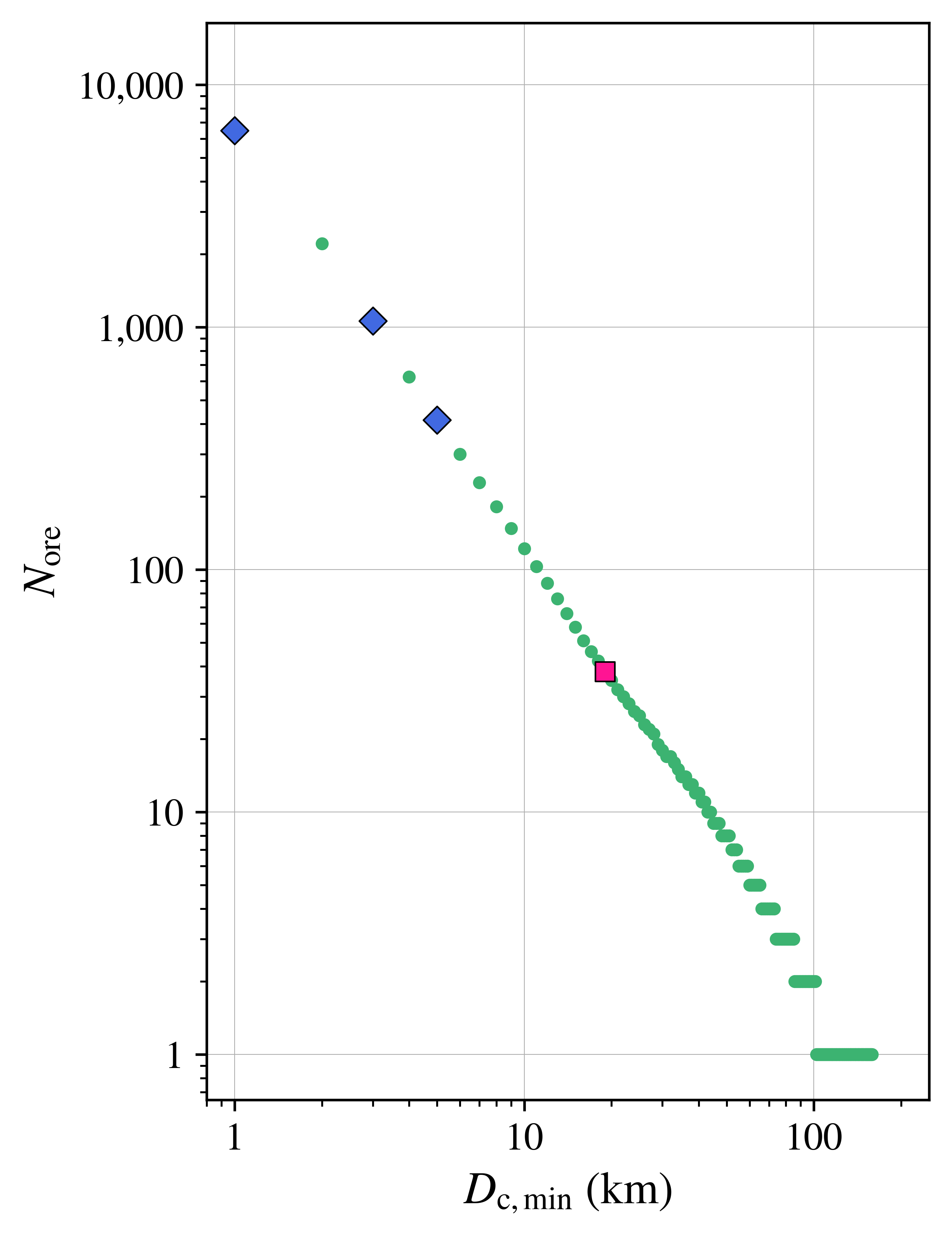}
    \caption{The number of PGM ore-bearing lunar craters as a function of crater diameter threshold. The three blue diamonds indicate 1\,km, 3\,km, and 5\,km. The pink square indicates the simple-to-complex transition diameter of 19\,km.}\label{fig:n_ore_vs_d_cmin_pgm}
\end{figure}

To examine how $N_{\rm ore}$ varies as a function of $D_{\rm c,min}$, we use the \citet{robbins2018} database. We use the crater diameter estimated based on a circular fit, and for each sample value of diameter, we calculate $N_{\rm ore}$. Fig.~\ref{fig:n_ore_vs_d_cmin_pgm} shows the resulting plot.

\begin{table*}[t]
    \centering
    \begin{tabular}{lrrrrrrr}
	\hline
	& $P_{\rm type}$ & $P_{\rm rich}$ & $P_{\rm surv}$ & $P_{\rm eng}$ & $D_{\rm c,min}$ & $N(> D_{\rm c,min})$ & $N_{\rm ore}$\\
	\hline
        PGMs & 0.04 & 0.5 & 0.25 & 1 & 1\,km & 1,296,796 & 6,484\\
        & & & & & 3\,km & 212,149 & 1,061\\
        & & & & & 5\,km & 83,061 & 415\\
        & & & & & 19\,km & 7,588 & 38\\
        Water & 0.1 & 0.31 & 0.083 & 1 & 1\,km & 1,296,796 & 3,350\\
        & & & & & 19\,km & 7,588 & 20\\
	\hline
    \end{tabular}
    \caption{Values of the terms in Eq.~(\ref{eq:mod_elvis}) and estimates of $N_{\rm ore}$ for PGM ore- and water-bearing lunar craters. Note that the values of $N_{\rm ore}$ are upper limits.}\label{tab:estimates}
\end{table*}

\subsection{Water}
\label{sec:water}

\citet{elvis2014} applies Eq.~(\ref{eq:elvis}) to C-type NEAs to estimate the number of water-bearing asteroids, assuming $P_{\rm type} = 0.1$ and $P_{\rm rich} = 0.31$\footnote{There is a discrepancy in this value in \citet{elvis2014}. On their page 24, they derive a value of $P_{\rm rich} = 0.31$ based on the results of \citet{jarosewich1990}, and use this in the calculation of $N_{\rm ore}$. However, elsewhere on the same page and in Table 2, they quote a value of 0.25. We adopt $P_{\rm rich} = 0.31$ in this paper.}. Motivated by detection considerations and ore value, they consider two cases regarding asteroid diameter thresholds: asteroids with $D_{\rm min} = 100$\,m whose absolute magnitude $H < 22$, and $D_{\rm min} = 18$\,m with $H > 22$. The quantity of ore in the smallest asteroids of the latter population is adequate for profitability, assuming a concentration of 20\% water by weight and a price of US\$5,000/kg for water in low-Earth orbit. However, smaller asteroids are harder to detect and characterize. For the larger threshold, $P_{\rm acc} = 0.025$, and for the smaller one, $P_{\rm acc} = 0.03$, since the distributions of delta-v values needed to reach them are different for the two populations \citep{elvis2011}. The number of NEAs larger than the threshold, $N(> D_{\rm min})$, is taken to be $2 \times 10^4$ for $D_{\rm min} = 100$\,m and $10^7$ for $D_{\rm min} = 18$\,m. $P_{\rm eng}$ is assumed to be unity as in the case of PGM ore-bearing asteroids. For these values, they calculate 18 NEAs $\ge 100$\,m in diameter\footnote{There is an error in their calculation of $N_{\rm ore}$ for this population. Using $P_{\rm acc} = 0.025$ and $N(> D_{\rm min}) = 2 \times 10^4$ yields $N_{\rm ore} = 16$. Since the difference is slight, we continue to quote the value of 18.} and $\sim$9000 NEAs $\ge 18$\,m in diameter.

As before, we adopt the \citet{elvis2014} values of $P_{\rm type}$ and $P_{\rm rich}$
for Eq.~(\ref{eq:mod_elvis}). In Section~\ref{sec:pgms} we take $P_{\rm surv} = 0.25$. For the case of water, however, it is important to also account for the fact that impact heating can significantly dehydrate the impacting body. Planar shock experiments on CM chondrites \citep{tyburczy1986}, demonstrate that impacts ranging from ${\sim}13$ to ${\sim}50$\,GPa (corresponding to velocities between ${\sim}1$ and ${\sim}2$\,km\,s$^{-1}$) can cause the loss of between ${\sim}14$\% and ${\sim}89$\% of an asteroid's water content. More recent hypervelocity impact experiments performed by \citet{daly2018} suggest that up to 30\% of water in projectiles can be retained in impact melt and remnants. \citet{king2021} estimate that heated CM chondrites have lost ${\sim}15$\% to $>65$\% of the water they once contained and find that this level of dehydration indicates shock pressures of ${\sim}20$ to ${\sim}50$\,GPa. To better reflect the water loss on impact and the probability of hydrated minerals surviving in the impact crater, therefore, we reduce $P_{\rm surv}$ by a factor of three, to 0.083.

We only consider larger asteroids, i.e., $D_{\rm min} = 100$\,m, which we nominally associate with a crater diameter threshold $D_{\rm c,min} = 1$\,km. There are 1,296,796 craters with diameters equal to or larger than this threshold. For these values of the terms, we obtain $N_{\rm ore} = 3,350$. We do not consider the smaller impactor diameter threshold, since closer to this threshold, asteroids predominantly create craters smaller than 1--2\,km, the census of which is incomplete \citep{robbins2018}.

\begin{figure}[t]
    \centering\includegraphics[width=0.5\linewidth, alt={Log-log plot with crater diameter threshold on the X-axis and the number of ore-bearing craters on the Y-axis. The points lie roughly along a line with a slope of approximately -2.}]{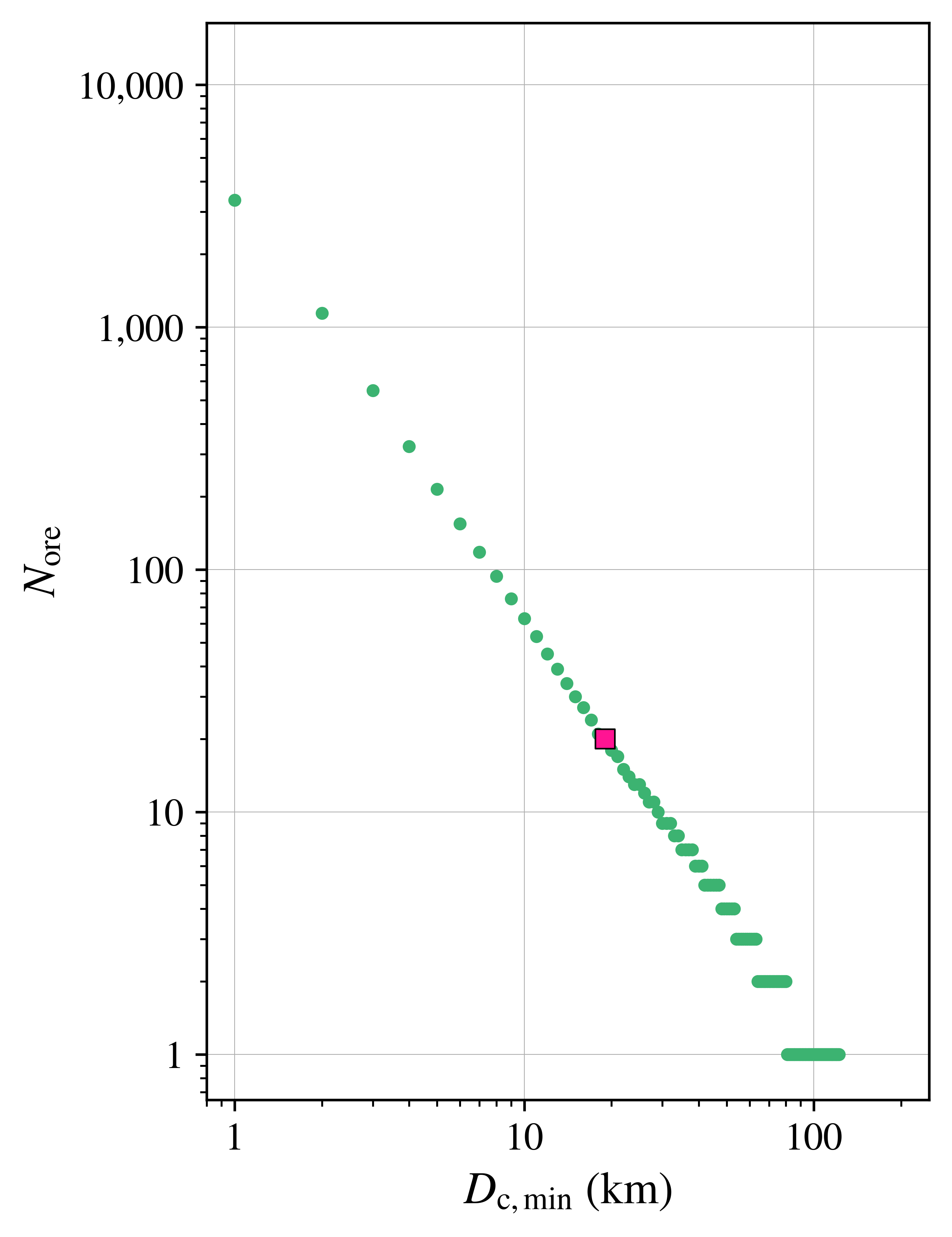}
    \caption{The number of water-bearing lunar craters as a function of crater diameter threshold. The pink square indicates the simple-to-complex transition diameter of 19\,km.}\label{fig:n_ore_vs_d_cmin_water}
\end{figure}

Since even the remnants of asteroids much smaller than 100\,m are profitable to mine for water, we explicitly report the number of craters for $D_{\rm c,min} = 1$\,km but not for 3\,km and 5\,km as we have done in Section~\ref{sec:pgms}. If we consider only complex craters, where the asteroid material is swept into the central feature instead of being spread throughout the breccia lens, and take the simple-to-complex transition diameter of $19$\,km \citep{kruger2018} as $D_{\rm c,min}$, we obtain $N_{\rm ore} = 20$. Fig.~\ref{fig:n_ore_vs_d_cmin_water} shows $N_{\rm ore}$ as a function of $D_{\rm c,min}$. Table~\ref{tab:estimates} summarizes the results.

\section{Discussion}
\label{sec:discussion}

\citet{elvis2014} obtains an $N_{\rm ore}$ of 10 for PGM ore-bearing NEAs and 18 for water-bearing ones. Depending on the choice of minimum crater diameter, our results for lunar craters are one to two orders of magnitude greater than that of NEAs. However, the estimates of $N_{\rm ore}$ for NEAs have changed over the last decade; since the publication of \citet{elvis2014} the number of known NEAs has tripled\footnote{\url{https://cneos.jpl.nasa.gov/stats/totals.html}; last accessed: 2024-08-10}, and the estimate of $N(> D_{\rm min})$ for $D_{\rm min} = 100$\,m has increased by 50\%, to $3 \times 10^4$ \citep{nesvorny2024}. New surveys from the Rubin Observatory \citep{jones2018} and NEO Surveyor\footnote{\url{https://www.jpl.nasa.gov/missions/near-earth-object-surveyor}; last accessed: 2024-08-10} will detect many more NEAs in the next few years. As a result, NEA population estimates may again be revised upwards. Additionally, upcoming advances in rocketry\footnote{\raggedright{\url{https://spaceflightnow.com/2024/06/06/live-coverage-spacex-to-launch-its-starship-rocket-on-its-fourth-test-flight/}}; last accessed: 2024-08-10} will increase the number of asteroids that we can realistically reach. Raising the delta-v threshold from $4.5$\,km\,s$^{-1}$ to $7$\,km\,s$^{-1}$ changes $P_{\rm acc}$ from 0.025 to 0.68 and results in a factor of ${\sim}27$ increase in $N_{\rm ore}$ \citep[see Fig. 1 of][]{elvis2014}. These developments would make $N_{\rm ore}$ for NEAs and lunar craters more comparable.

Following \citet{elvis2014}, we have set $P_{\rm eng} = 1$ in our analysis, yielding upper limits for $N_{\rm ore}$. The true value of this term is challenging to assess as it depends on several factors, with the concentration of ore within the lunar regolith being particularly important. This concentration, by definition, depends on two key factors: (i) the size of the ore fragments, and (ii) the degree to which these fragments are dispersed throughout the impact site. Both of these factors are, in turn, affected by the impact conditions, such as the angle and velocity of the impactor upon collision, as well as its size. These parameters also determine whether the resulting crater is simple or complex in structure. As shown by \citet{yue2013}, during the formation of complex craters in vertical impacts, the projectile material is initially dispersed, but is collected in the central peak during the collapse of the crater. This is different from a simple crater, in which the material remains dispersed within the breccia lens. In general, if the resulting impactor fragments are small and highly dispersed across the regolith, the material may not be viable for mining due to the increased difficulty in extraction. Conversely, larger fragments or concentrated regions of ore would present a more favorable scenario for mining operations. Because of the complications outlined above, we limit ourselves to a high-level comparison of $P_{\rm eng}$ when prospecting for and mining resources in lunar asteroid remnants versus NEAs in orbit. The advantages of mining asteroid remnants on the Moon include: (i) the relative ease of prospecting due to its proximity, (ii) a relatively high surface gravity, typically thousands of times larger than that of an asteroid, (iii) the feasibility of teleoperation of machines from Earth due to its proximity, (iv) the possibility of human presence, which makes mining activities more productive than if carried out entirely by robots, and (v) a higher surface stability than  asteroids, which rotate or tumble. Conversely, the advantages of mining NEAs include: (i) a higher concentration of resources due to the lack of additional regolith mixing, and (ii) the existence of a population of asteroids with a lower delta-v than the Moon, making this subset of cases more cost-effective to access. The relative merits, together with our estimates of $N_{\rm ore}$, \emph{prima facie} suggest that the Moon may be a more advantageous, and hence more profitable, target for asteroid mining endeavors than NEAs.

Several caveats need to be considered with regard to our analysis. Firstly, in Eq.~(\ref{eq:elvis}) and Eq.~(\ref{eq:mod_elvis}) there is an interdependence between the probability of resource richness $P_{\text{rich}}$ and the number of NEOs larger than a minimum profitable mass $N(> M_{\text{min}})$. A higher $P_{\text{rich}}$ allows for a lower threshold for $N(> M_{\text{min}})$, as even smaller asteroids can contribute significantly to profitability due to their richness in the resource. Conversely, a lower $P_{\text{rich}}$ necessitates a higher threshold for $N(> M_{\text{min}})$ to ensure that profitability is achieved from fewer, larger asteroids. This relationship should be considered when setting thresholds.

The values of $P_{\rm type}$ and $P_{\rm rich}$ that we use in our calculations are based on the NEA population that we observe today. The impact flux on the Moon has changed over the past ${\sim}4.5$\,Gyr \citep[see, e.g.,][]{bottke2012}, and it is possible that the contributions of M- and C-type asteroids to the impact flux has also evolved over time. This can change our estimates of $N_{\rm ore}$. Additionally, the craters on the Moon were formed not just by asteroids but also by comets, although the contribution of comets may have only been ${\sim} 1\%$ \citep{nesvorny2023}.

There are limitations to the estimate of $P_{\rm surv}$. The simulations performed by \citet{yue2013} assume dunite (an igneous rock consisting largely of olivine) as the impactor material and granite as the target material, and \cite{svetsov2015} assume both dunite and quartz impactors on quartz and gabbroic anorthosite targets. Therefore, these simulations are approximations of the actual compositions of impacting meteorites and the lunar surface, which can be more complex and heterogeneous. However, \citet{svetsov2015} note that the choice of the impactor and target materials on the remnant mass is insignificant. The \citet{yue2013} simulation was two-dimensional, and the impactor was 7\,km in diameter, resulting in a complex crater with a rim-to-rim diameter of ${\sim}89$\,km. The \citet{svetsov2015} simulations used 1\,km-diameter projectiles. They note that the size of the asteroid may have an effect on the fraction of material remaining in the crater. Additionally, the porosity of the asteroid and the regolith may also affect how much of the impactor survives \citep[see, e.g.,][]{kalynn2013}. In the case of water-bearing asteroids, $P_{\rm surv}$ is also affected by impact heating as discussed in Section~\ref{sec:water}, which has the effect of decreasing the amount of remaining water content. As discussed in Section~\ref{sec:pgms}, the impact angle will also determine the fraction of material that remains in the crater.

In the vertical-impact simulations of \citet[][]{yue2013}, they found that impactor material gets concentrated in the central features of complex craters. There is scant observational evidence for this, however, most likely because (a) vertical impacts are rare, and (b) determining the impact angle is difficult, so identifying such craters is a challenge. There have been several studies of lunar crater central peaks \cite[see, e.g.,][]{song2013, dhingra2015}. \citet{dhingra2015} discuss the origin of olivine in the central peak of the crater Copernicus and report that they find no evidence of an exogenic origin. However, while not a complex crater with a central peak, \citet[][]{yang2022} report the detection of carbonaceous chondrite remnants at the center of a $\sim$2\,m-diameter crater, but note that it could have been produced by a secondary impact. \citet[][]{klima2017} note that distinguishing between endogenic and exogenic hydroxyls is difficult via remote sensing, so in the case of water-bearing craters, in situ measurements may be required. Detecting PGM ore-bearing remnants in central peaks requires sensors with the ability to detect metallic iron or nickel at a spatial resolution of ${\sim}1$ to ${\sim}10$\,km, and no such sensors have been flown hitherto. In any case, the accumulation of asteroidal material into central features of complex craters requires vertical or near-vertical impacts, but most impacts are oblique, which result in debris spread elsewhere in the crater or down range.

The \citet{robbins2018} database is complete for craters larger than 1--2\,km, with the number of missed craters increasing towards the lower bound of that range. A more complete lunar crater census, especially for crater diameters less than 2\,km will help refine our results. In addition, better estimates of progenitor asteroid sizes from crater sizes will help constrain $N_{\rm ore}$. More sophisticated simulations, e.g., three-dimensional simulations of more representative materials, including those that make up M- and C-type asteroids, and simulations that take into account the distributions of asteroid sizes, densities, impact velocities, and impact angles, will help determine the amount of asteroidal material that survives in lunar craters.

For water-bearing asteroids, additional caveats need to be considered. Determining the fraction of NEAs that are hydrated is challenging, as discussed in \citet{rivkin2019}. Based on the method of derivation, they consider values of $P_{\rm rich}$ ranging from 0.02 to 0.19. These are well short of the value of 0.31 that we took from \citet{elvis2014}. If the true hydrated asteroid fraction is closer to these lower numbers, then the estimates of ore-bearing craters will be commensurately lower. With regard to water content within hydrated asteroids, \citet{elvis2014} assumes a concentration of 20\% water by weight in C-type asteroids. Of this, half is assumed to be present in the form of ice, which would vaporize on impact, reducing the ore value. Additionally, the assumption of US\$5,000/kg for water is made with regard to low-Earth orbit. The value of water will likely be different on the Moon. Water, in the form of ice, is believed to be present in the permanently shadowed areas of the lunar south polar region \citep{colaprete2010}, so asteroidal water may not be valuable there. Away from the poles, though, asteroidal water present in hydrated minerals may be more valuable.

The argument given in this paper is statistical and does not identify which craters are ore-rich. Validating our results and identifying PGM-rich craters require a significant prospecting effort. Appropriate instrumentation could potentially be integrated into upcoming lunar missions to prospect for metallic iron or nickel, proxies for PGMs. Prospecting could be achieved using remote sensing spacecraft in lunar orbit or landers making in situ measurements. In the case of remote sensing, craters that are unresolved spatially will have their apparent iron and nickel content diluted by their low-concentration surroundings. Hence, a spatial resolution of $\lesssim1$\,km for the large number of small craters, or $\lesssim10$\,km for the ${\sim}40$ largest ones is required. Recall from Table~\ref{tab:estimates} that ${\sim}7,500$ complex craters need to be mapped to find the ${\sim}40$ best ore-bodies. Therefore, with a hit rate of roughly 0.5\%, in situ investigations would seem to be a poor approach. Remote sensing from lunar orbit is far more preferable, if feasible. However, characterization of mineable materials and determining how they might be extracted from the regolith would require in situ studies.

Since the total mass of PGMs in ore-rich craters increases roughly as the cube of crater diameter, the fewer large craters are far more desirable mining sites than the more numerous smaller ones. Other lunar resources are similarly concentrated in a few dozen locations that are typically a few kilometers in diameter \citep{elvis2020}. This situation is one in which multiple players will wish to access the same sites, with a concomitant risk of conflict. Asteroid debris sites now add to that problematic situation.

In summary, our analysis suggests a number of lunar craters containing ore-bearing asteroid remnants that is higher than the number of currently-accessible ore-bearing NEAs and so highlights the potential viability and profitability of lunar mining endeavours compared to mining asteroids in orbit.

\section*{Acknowledgments}

We thank William Bottke, Brandon Johnson, Kai W\"unnemann, and David Lawrence for useful discussions. We also thank the anonymous reviewers for comments that served to clarify our arguments.

\bibliographystyle{elsarticle-harv} 
\bibliography{references}

\begin{thebibliography}{31}
\expandafter\ifx\csname natexlab\endcsname\relax\def\natexlab#1{#1}\fi
\providecommand{\url}[1]{\texttt{#1}}
\providecommand{\href}[2]{#2}
\providecommand{\path}[1]{#1}
\providecommand{\DOIprefix}{doi:}
\providecommand{\ArXivprefix}{arXiv:}
\providecommand{\URLprefix}{URL: }
\providecommand{\Pubmedprefix}{pmid:}
\providecommand{\doi}[1]{\href{http://dx.doi.org/#1}{\path{#1}}}
\providecommand{\Pubmed}[1]{\href{pmid:#1}{\path{#1}}}
\providecommand{\bibinfo}[2]{#2}
\ifx\xfnm\relax \def\xfnm[#1]{\unskip,\space#1}\fi
\bibitem[{{Bland} et~al.(2008){Bland}, {Artemieva}, {Collins}, {Bottke}, {Bussey} and {Joy}}]{bland2008}
\bibinfo{author}{{Bland}, P.A.}, \bibinfo{author}{{Artemieva}, N.A.}, \bibinfo{author}{{Collins}, G.S.}, \bibinfo{author}{{Bottke}, W.F.}, \bibinfo{author}{{Bussey}, D.B.J.}, \bibinfo{author}{{Joy}, K.H.}, \bibinfo{year}{2008}.
\newblock \bibinfo{title}{{Asteroids on the Moon: Projectile Survival During Low Velocity Impact}}, in: \bibinfo{booktitle}{39th Annual Lunar and Planetary Science Conference}, p. \bibinfo{pages}{2045}.
\bibitem[{{Bottke} et~al.(2012){Bottke}, {Vokrouhlick{\'y}}, {Minton}, {Nesvorn{\'y}}, {Morbidelli}, {Brasser}, {Simonson} and {Levison}}]{bottke2012}
\bibinfo{author}{{Bottke}, W.F.}, \bibinfo{author}{{Vokrouhlick{\'y}}, D.}, \bibinfo{author}{{Minton}, D.}, \bibinfo{author}{{Nesvorn{\'y}}, D.}, \bibinfo{author}{{Morbidelli}, A.}, \bibinfo{author}{{Brasser}, R.}, \bibinfo{author}{{Simonson}, B.}, \bibinfo{author}{{Levison}, H.F.}, \bibinfo{year}{2012}.
\newblock \bibinfo{title}{{An Archaean heavy bombardment from a destabilized extension of the asteroid belt}}.
\newblock \bibinfo{journal}{Nature} \bibinfo{volume}{485}, \bibinfo{pages}{78--81}.
\newblock \DOIprefix\doi{10.1038/nature10967}.
\bibitem[{{Colaprete} et~al.(2010){Colaprete}, {Schultz}, {Heldmann}, {Wooden}, {Shirley}, {Ennico}, {Hermalyn}, {Marshall}, {Ricco}, {Elphic}, {Goldstein}, {Summy}, {Bart}, {Asphaug}, {Korycansky}, {Landis} and {Sollitt}}]{colaprete2010}
\bibinfo{author}{{Colaprete}, A.}, \bibinfo{author}{{Schultz}, P.}, \bibinfo{author}{{Heldmann}, J.}, \bibinfo{author}{{Wooden}, D.}, \bibinfo{author}{{Shirley}, M.}, \bibinfo{author}{{Ennico}, K.}, \bibinfo{author}{{Hermalyn}, B.}, \bibinfo{author}{{Marshall}, W.}, \bibinfo{author}{{Ricco}, A.}, \bibinfo{author}{{Elphic}, R.C.}, \bibinfo{author}{{Goldstein}, D.}, \bibinfo{author}{{Summy}, D.}, \bibinfo{author}{{Bart}, G.D.}, \bibinfo{author}{{Asphaug}, E.}, \bibinfo{author}{{Korycansky}, D.}, \bibinfo{author}{{Landis}, D.}, \bibinfo{author}{{Sollitt}, L.}, \bibinfo{year}{2010}.
\newblock \bibinfo{title}{{Detection of Water in the LCROSS Ejecta Plume}}.
\newblock \bibinfo{journal}{Science} \bibinfo{volume}{330}, \bibinfo{pages}{463}.
\newblock \DOIprefix\doi{10.1126/science.1186986}.
\bibitem[{{Collins} et~al.(2005){Collins}, {Melosh} and {Marcus}}]{collins2005}
\bibinfo{author}{{Collins}, G.S.}, \bibinfo{author}{{Melosh}, H.J.}, \bibinfo{author}{{Marcus}, R.A.}, \bibinfo{year}{2005}.
\newblock \bibinfo{title}{{Earth Impact Effects Program: A Web-based computer program for calculating the regional environmental consequences of a meteoroid impact on Earth}}.
\newblock \bibinfo{journal}{Meteoritics and Planetary Science} \bibinfo{volume}{40}, \bibinfo{pages}{817}.
\newblock \DOIprefix\doi{10.1111/j.1945-5100.2005.tb00157.x}.
\bibitem[{{Crawford} et~al.(2023){Crawford}, {Anand}, {Barber}, {Cowley}, {Crites}, {Fa}, {Flahaut}, {Gaddis}, {Greenhagen}, {Haruyama}, {Hurley}, {McLeod}, {Morse}, {Neal}, {Sargeant}, {Sefton-Nash} and {Tart{\`e}se}}]{crawford2023}
\bibinfo{author}{{Crawford}, I.A.}, \bibinfo{author}{{Anand}, M.}, \bibinfo{author}{{Barber}, S.}, \bibinfo{author}{{Cowley}, A.}, \bibinfo{author}{{Crites}, S.}, \bibinfo{author}{{Fa}, W.}, \bibinfo{author}{{Flahaut}, J.}, \bibinfo{author}{{Gaddis}, L.R.}, \bibinfo{author}{{Greenhagen}, B.}, \bibinfo{author}{{Haruyama}, J.}, \bibinfo{author}{{Hurley}, D.}, \bibinfo{author}{{McLeod}, C.L.}, \bibinfo{author}{{Morse}, A.}, \bibinfo{author}{{Neal}, C.R.}, \bibinfo{author}{{Sargeant}, H.}, \bibinfo{author}{{Sefton-Nash}, E.}, \bibinfo{author}{{Tart{\`e}se}, R.}, \bibinfo{year}{2023}.
\newblock \bibinfo{title}{{Lunar Resources}}.
\newblock \bibinfo{journal}{Reviews in Mineralogy and Geochemistry} \bibinfo{volume}{89}, \bibinfo{pages}{829--868}.
\newblock \DOIprefix\doi{10.2138/rmg.2023.89.19}.
\bibitem[{Daly and Schultz(2018)}]{daly2018}
\bibinfo{author}{Daly, R.T.}, \bibinfo{author}{Schultz, P.H.}, \bibinfo{year}{2018}.
\newblock \bibinfo{title}{The delivery of water by impacts from planetary accretion to present}.
\newblock \bibinfo{journal}{Science Advances} \bibinfo{volume}{4}, \bibinfo{pages}{eaar2632}.
\newblock \URLprefix \url{https://www.science.org/doi/abs/10.1126/sciadv.aar2632}, \DOIprefix\doi{10.1126/sciadv.aar2632}, \href{http://arxiv.org/abs/https://www.science.org/doi/pdf/10.1126/sciadv.aar2632}{{\tt arXiv:https://www.science.org/doi/pdf/10.1126/sciadv.aar2632}}.
\bibitem[{{Dhingra} et~al.(2015){Dhingra}, {Pieters} and {Head}}]{dhingra2015}
\bibinfo{author}{{Dhingra}, D.}, \bibinfo{author}{{Pieters}, C.M.}, \bibinfo{author}{{Head}, J.W.}, \bibinfo{year}{2015}.
\newblock \bibinfo{title}{{Multiple origins for olivine at Copernicus crater}}.
\newblock \bibinfo{journal}{Earth and Planetary Science Letters} \bibinfo{volume}{420}, \bibinfo{pages}{95--101}.
\newblock \DOIprefix\doi{10.1016/j.epsl.2015.03.039}.
\bibitem[{{Elvis}(2014)}]{elvis2014}
\bibinfo{author}{{Elvis}, M.}, \bibinfo{year}{2014}.
\newblock \bibinfo{title}{{How many ore-bearing asteroids?}}
\newblock \bibinfo{journal}{Planetary and Space Science} \bibinfo{volume}{91}, \bibinfo{pages}{20--26}.
\newblock \DOIprefix\doi{10.1016/j.pss.2013.11.008}, \href{http://arxiv.org/abs/1312.4450}{{\tt arXiv:1312.4450}}.
\bibitem[{{Elvis} et~al.(2020){Elvis}, {Krolikowski} and {Milligan}}]{elvis2020}
\bibinfo{author}{{Elvis}, M.}, \bibinfo{author}{{Krolikowski}, A.}, \bibinfo{author}{{Milligan}, T.}, \bibinfo{year}{2020}.
\newblock \bibinfo{title}{{Concentrated lunar resources: imminent implications for governance and justice}}.
\newblock \bibinfo{journal}{Philosophical Transactions of the Royal Society of London Series A} \bibinfo{volume}{379}, \bibinfo{pages}{20190563}.
\newblock \DOIprefix\doi{10.1098/rsta.2019.0563}, \href{http://arxiv.org/abs/2103.09045}{{\tt arXiv:2103.09045}}.
\bibitem[{{Elvis} et~al.(2011){Elvis}, {McDowell}, {Hoffman} and {Binzel}}]{elvis2011}
\bibinfo{author}{{Elvis}, M.}, \bibinfo{author}{{McDowell}, J.}, \bibinfo{author}{{Hoffman}, J.A.}, \bibinfo{author}{{Binzel}, R.P.}, \bibinfo{year}{2011}.
\newblock \bibinfo{title}{{Ultra-low delta-v objects and the human exploration of asteroids}}.
\newblock \bibinfo{journal}{Planetary and Space Science} \bibinfo{volume}{59}, \bibinfo{pages}{1408--1412}.
\newblock \DOIprefix\doi{10.1016/j.pss.2011.05.006}, \href{http://arxiv.org/abs/1105.4152}{{\tt arXiv:1105.4152}}.
\bibitem[{{Halim} et~al.(2024){Halim}, {Crawford}, {Collins}, {Joy} and {Davison}}]{halim2024}
\bibinfo{author}{{Halim}, S.H.}, \bibinfo{author}{{Crawford}, I.A.}, \bibinfo{author}{{Collins}, G.S.}, \bibinfo{author}{{Joy}, K.H.}, \bibinfo{author}{{Davison}, T.M.}, \bibinfo{year}{2024}.
\newblock \bibinfo{title}{{Assessing the survival of carbonaceous chondrites impacting the lunar surface as a potential resource}}.
\newblock \bibinfo{journal}{Planetary and Space Science} \bibinfo{volume}{246}, \bibinfo{pages}{105905}.
\newblock \DOIprefix\doi{10.1016/j.pss.2024.105905}.
\bibitem[{{Jarosewich}(1990)}]{jarosewich1990}
\bibinfo{author}{{Jarosewich}, E.}, \bibinfo{year}{1990}.
\newblock \bibinfo{title}{{Chemical Analyses of Meteorites: A Compilation of Stony and Iron Meteorite Analyses}}.
\newblock \bibinfo{journal}{Meteoritics} \bibinfo{volume}{25}, \bibinfo{pages}{323}.
\newblock \DOIprefix\doi{10.1111/j.1945-5100.1990.tb00717.x}.
\bibitem[{{Johnson} et~al.(2016){Johnson}, {Collins}, {Minton}, {Bowling}, {Simonson} and {Zuber}}]{johnson2016}
\bibinfo{author}{{Johnson}, B.C.}, \bibinfo{author}{{Collins}, G.S.}, \bibinfo{author}{{Minton}, D.A.}, \bibinfo{author}{{Bowling}, T.J.}, \bibinfo{author}{{Simonson}, B.M.}, \bibinfo{author}{{Zuber}, M.T.}, \bibinfo{year}{2016}.
\newblock \bibinfo{title}{{Spherule layers, crater scaling laws, and the population of ancient terrestrial impactors}}.
\newblock \bibinfo{journal}{Icarus} \bibinfo{volume}{271}, \bibinfo{pages}{350--359}.
\newblock \DOIprefix\doi{10.1016/j.icarus.2016.02.023}.
\bibitem[{{Jones} et~al.(2018){Jones}, {Slater}, {Moeyens}, {Allen}, {Axelrod}, {Cook}, {Ivezi{\'c}}, {Juri{\'c}}, {Myers} and {Petry}}]{jones2018}
\bibinfo{author}{{Jones}, R.L.}, \bibinfo{author}{{Slater}, C.T.}, \bibinfo{author}{{Moeyens}, J.}, \bibinfo{author}{{Allen}, L.}, \bibinfo{author}{{Axelrod}, T.}, \bibinfo{author}{{Cook}, K.}, \bibinfo{author}{{Ivezi{\'c}}, {\v{Z}}.}, \bibinfo{author}{{Juri{\'c}}, M.}, \bibinfo{author}{{Myers}, J.}, \bibinfo{author}{{Petry}, C.E.}, \bibinfo{year}{2018}.
\newblock \bibinfo{title}{{The Large Synoptic Survey Telescope as a Near-Earth Object discovery machine}}.
\newblock \bibinfo{journal}{Icarus} \bibinfo{volume}{303}, \bibinfo{pages}{181--202}.
\newblock \DOIprefix\doi{10.1016/j.icarus.2017.11.033}, \href{http://arxiv.org/abs/1711.10621}{{\tt arXiv:1711.10621}}.
\bibitem[{{Kalynn} et~al.(2013){Kalynn}, {Johnson}, {Osinski} and {Barnouin}}]{kalynn2013}
\bibinfo{author}{{Kalynn}, J.}, \bibinfo{author}{{Johnson}, C.L.}, \bibinfo{author}{{Osinski}, G.R.}, \bibinfo{author}{{Barnouin}, O.}, \bibinfo{year}{2013}.
\newblock \bibinfo{title}{{Topographic characterization of lunar complex craters}}.
\newblock \bibinfo{journal}{Geophysical Research Letters} \bibinfo{volume}{40}, \bibinfo{pages}{38--42}.
\newblock \DOIprefix\doi{10.1029/2012GL053608}.
\bibitem[{{King} et~al.(2021){King}, {Schofield} and {Russell}}]{king2021}
\bibinfo{author}{{King}, A.J.}, \bibinfo{author}{{Schofield}, P.F.}, \bibinfo{author}{{Russell}, S.S.}, \bibinfo{year}{2021}.
\newblock \bibinfo{title}{{Thermal alteration of CM carbonaceous chondrites: Mineralogical changes and metamorphic temperatures}}.
\newblock \bibinfo{journal}{Geochimica et Cosmochimica Acta} \bibinfo{volume}{298}, \bibinfo{pages}{167--190}.
\newblock \DOIprefix\doi{10.1016/j.gca.2021.02.011}, \href{http://arxiv.org/abs/2102.07634}{{\tt arXiv:2102.07634}}.
\bibitem[{{Klima} and {Petro}(2017)}]{klima2017}
\bibinfo{author}{{Klima}, R.L.}, \bibinfo{author}{{Petro}, N.E.}, \bibinfo{year}{2017}.
\newblock \bibinfo{title}{{Remotely distinguishing and mapping endogenic water on the Moon}}.
\newblock \bibinfo{journal}{Philosophical Transactions of the Royal Society of London Series A} \bibinfo{volume}{375}, \bibinfo{pages}{20150391}.
\newblock \DOIprefix\doi{10.1098/rsta.2015.0391}.
\bibitem[{{Kr{\"u}ger} et~al.(2018){Kr{\"u}ger}, {Hergarten} and {Kenkmann}}]{kruger2018}
\bibinfo{author}{{Kr{\"u}ger}, T.}, \bibinfo{author}{{Hergarten}, S.}, \bibinfo{author}{{Kenkmann}, T.}, \bibinfo{year}{2018}.
\newblock \bibinfo{title}{{Deriving Morphometric Parameters and the Simple-to-Complex Transition Diameter From a High-Resolution, Global Database of Fresh Lunar Impact Craters (D {\ensuremath{\geq}} 3 km)}}.
\newblock \bibinfo{journal}{Journal of Geophysical Research (Planets)} \bibinfo{volume}{123}, \bibinfo{pages}{2667--2690}.
\newblock \DOIprefix\doi{10.1029/2018JE005545}.
\bibitem[{{Lewis}(1996)}]{lewis1996}
\bibinfo{author}{{Lewis}, J.S.}, \bibinfo{year}{1996}.
\newblock \bibinfo{title}{{Mining the Sky: Untold Riches from the Asteroids, Comets, and Planets}}.
\newblock \bibinfo{publisher}{Helix Books}.
\bibitem[{{Marchi} et~al.(2009){Marchi}, {Mottola}, {Cremonese}, {Massironi} and {Martellato}}]{marchi2009}
\bibinfo{author}{{Marchi}, S.}, \bibinfo{author}{{Mottola}, S.}, \bibinfo{author}{{Cremonese}, G.}, \bibinfo{author}{{Massironi}, M.}, \bibinfo{author}{{Martellato}, E.}, \bibinfo{year}{2009}.
\newblock \bibinfo{title}{{A New Chronology for the Moon and Mercury}}.
\newblock \bibinfo{journal}{The Astronomical Journal} \bibinfo{volume}{137}, \bibinfo{pages}{4936--4948}.
\newblock \DOIprefix\doi{10.1088/0004-6256/137/6/4936}, \href{http://arxiv.org/abs/0903.5137}{{\tt arXiv:0903.5137}}.
\bibitem[{{Nesvorn{\'y}} et~al.(2023){Nesvorn{\'y}}, {Deienno}, {Bottke}, {Jedicke}, {Naidu}, {Chesley}, {Chodas}, {Granvik}, {Vokrouhlick{\'y}}, {Bro{\v{z}}}, {Morbidelli}, {Christensen}, {Shelly} and {Bolin}}]{nesvorny2023}
\bibinfo{author}{{Nesvorn{\'y}}, D.}, \bibinfo{author}{{Deienno}, R.}, \bibinfo{author}{{Bottke}, W.F.}, \bibinfo{author}{{Jedicke}, R.}, \bibinfo{author}{{Naidu}, S.}, \bibinfo{author}{{Chesley}, S.R.}, \bibinfo{author}{{Chodas}, P.W.}, \bibinfo{author}{{Granvik}, M.}, \bibinfo{author}{{Vokrouhlick{\'y}}, D.}, \bibinfo{author}{{Bro{\v{z}}}, M.}, \bibinfo{author}{{Morbidelli}, A.}, \bibinfo{author}{{Christensen}, E.}, \bibinfo{author}{{Shelly}, F.C.}, \bibinfo{author}{{Bolin}, B.T.}, \bibinfo{year}{2023}.
\newblock \bibinfo{title}{{NEOMOD: A New Orbital Distribution Model for Near-Earth Objects}}.
\newblock \bibinfo{journal}{The Astronomical Journal} \bibinfo{volume}{166}, \bibinfo{pages}{55}.
\newblock \DOIprefix\doi{10.3847/1538-3881/ace040}, \href{http://arxiv.org/abs/2306.09521}{{\tt arXiv:2306.09521}}.
\bibitem[{{Nesvorn{\'y}} et~al.(2024){Nesvorn{\'y}}, {Vokrouhlick{\'y}}, {Shelly}, {Deienno}, {Bottke}, {Fuls}, {Jedicke}, {Naidu}, {Chesley}, {Chodas}, {Farnocchia} and {Delbo}}]{nesvorny2024}
\bibinfo{author}{{Nesvorn{\'y}}, D.}, \bibinfo{author}{{Vokrouhlick{\'y}}, D.}, \bibinfo{author}{{Shelly}, F.}, \bibinfo{author}{{Deienno}, R.}, \bibinfo{author}{{Bottke}, W.F.}, \bibinfo{author}{{Fuls}, C.}, \bibinfo{author}{{Jedicke}, R.}, \bibinfo{author}{{Naidu}, S.}, \bibinfo{author}{{Chesley}, S.R.}, \bibinfo{author}{{Chodas}, P.W.}, \bibinfo{author}{{Farnocchia}, D.}, \bibinfo{author}{{Delbo}, M.}, \bibinfo{year}{2024}.
\newblock \bibinfo{title}{{NEOMOD 3: The debiased size distribution of Near Earth Objects}}.
\newblock \bibinfo{journal}{Icarus} \bibinfo{volume}{417}, \bibinfo{pages}{116110}.
\newblock \DOIprefix\doi{10.1016/j.icarus.2024.116110}, \href{http://arxiv.org/abs/2404.18805}{{\tt arXiv:2404.18805}}.
\bibitem[{{Rivkin} and {DeMeo}(2019)}]{rivkin2019}
\bibinfo{author}{{Rivkin}, A.S.}, \bibinfo{author}{{DeMeo}, F.E.}, \bibinfo{year}{2019}.
\newblock \bibinfo{title}{{How Many Hydrated NEOs Are There?}}
\newblock \bibinfo{journal}{Journal of Geophysical Research (Planets)} \bibinfo{volume}{124}, \bibinfo{pages}{128--142}.
\newblock \DOIprefix\doi{10.1029/2018JE005584}, \href{http://arxiv.org/abs/1812.02285}{{\tt arXiv:1812.02285}}.
\bibitem[{{Robbins}(2018)}]{robbins2018}
\bibinfo{author}{{Robbins}, S.J.}, \bibinfo{year}{2018}.
\newblock \bibinfo{title}{{A Global Lunar Crater Database, Complete for Craters {\ensuremath{\geq}}1 km, III: Reassessing the Lunar Crater Production Function, and Lessons Learned Applied to the Global Mars Crater Database}}, in: \bibinfo{booktitle}{49th Annual Lunar and Planetary Science Conference}, p. \bibinfo{pages}{2443}.
\bibitem[{{Shoemaker}(1961)}]{shoemaker1961}
\bibinfo{author}{{Shoemaker}, E.M.}, \bibinfo{year}{1961}.
\newblock \bibinfo{title}{{Interpretation of Lunar Craters}}, in: \bibinfo{booktitle}{Physics and Astronomy of the Moon}. \bibinfo{publisher}{Academic Press Inc.}, pp. \bibinfo{pages}{283--359}.
\newblock \DOIprefix\doi{10.1016/B978-1-4832-3240-9.50012-2}.
\bibitem[{{Song} et~al.(2013){Song}, {Bandfield}, {Lucey}, {Greenhagen} and {Paige}}]{song2013}
\bibinfo{author}{{Song}, E.}, \bibinfo{author}{{Bandfield}, J.L.}, \bibinfo{author}{{Lucey}, P.G.}, \bibinfo{author}{{Greenhagen}, B.T.}, \bibinfo{author}{{Paige}, D.A.}, \bibinfo{year}{2013}.
\newblock \bibinfo{title}{{Bulk mineralogy of lunar crater central peaks via thermal infrared spectra from the Diviner Lunar Radiometer: A study of the Moon's crustal composition at depth}}.
\newblock \bibinfo{journal}{Journal of Geophysical Research (Planets)} \bibinfo{volume}{118}, \bibinfo{pages}{689--707}.
\newblock \DOIprefix\doi{10.1002/jgre.20065}.
\bibitem[{{Svetsov} and {Shuvalov}(2015)}]{svetsov2015}
\bibinfo{author}{{Svetsov}, V.V.}, \bibinfo{author}{{Shuvalov}, V.V.}, \bibinfo{year}{2015}.
\newblock \bibinfo{title}{{Water delivery to the Moon by asteroidal and cometary impacts}}.
\newblock \bibinfo{journal}{Planetary and Space Science} \bibinfo{volume}{117}, \bibinfo{pages}{444--452}.
\newblock \DOIprefix\doi{10.1016/j.pss.2015.09.011}.
\bibitem[{Tyburczy et~al.(1986)Tyburczy, Frisch and Ahrens}]{tyburczy1986}
\bibinfo{author}{Tyburczy, J.A.}, \bibinfo{author}{Frisch, B.}, \bibinfo{author}{Ahrens, T.J.}, \bibinfo{year}{1986}.
\newblock \bibinfo{title}{Shock-induced volatile loss from a carbonaceous chondrite: implications for planetary accretion}.
\newblock \bibinfo{journal}{Earth and Planetary Science Letters} \bibinfo{volume}{80}, \bibinfo{pages}{201--207}.
\newblock \URLprefix \url{https://www.sciencedirect.com/science/article/pii/0012821X86901044}, \DOIprefix\doi{https://doi.org/10.1016/0012-821X(86)90104-4}.
\bibitem[{{Wingo}(2004)}]{wingo2004}
\bibinfo{author}{{Wingo}, D.}, \bibinfo{year}{2004}.
\newblock \bibinfo{title}{{Moonrush: Improving Life on Earth with the Moon's Resources}}.
\newblock \bibinfo{publisher}{Apogee Books}.
\bibitem[{{Yang} et~al.(2021){Yang}, {Li}, {Zhu}, {Liu}, {Wu}, {Du}, {Fa}, {Xu}, {He}, {Wang}, {Xue}, {Yang} and {Zou}}]{yang2022}
\bibinfo{author}{{Yang}, Y.}, \bibinfo{author}{{Li}, S.}, \bibinfo{author}{{Zhu}, M.H.}, \bibinfo{author}{{Liu}, Y.}, \bibinfo{author}{{Wu}, B.}, \bibinfo{author}{{Du}, J.}, \bibinfo{author}{{Fa}, W.}, \bibinfo{author}{{Xu}, R.}, \bibinfo{author}{{He}, Z.}, \bibinfo{author}{{Wang}, C.}, \bibinfo{author}{{Xue}, B.}, \bibinfo{author}{{Yang}, J.}, \bibinfo{author}{{Zou}, Y.}, \bibinfo{year}{2021}.
\newblock \bibinfo{title}{{Impact remnants rich in carbonaceous chondrites detected on the Moon by the Chang'e-4 rover}}.
\newblock \bibinfo{journal}{Nature Astronomy} \bibinfo{volume}{6}, \bibinfo{pages}{207--213}.
\newblock \DOIprefix\doi{10.1038/s41550-021-01530-w}.
\bibitem[{{Yue} et~al.(2013){Yue}, {Johnson}, {Minton}, {Melosh}, {di}, {Hu} and {Liu}}]{yue2013}
\bibinfo{author}{{Yue}, Z.}, \bibinfo{author}{{Johnson}, B.C.}, \bibinfo{author}{{Minton}, D.A.}, \bibinfo{author}{{Melosh}, H.J.}, \bibinfo{author}{{di}, K.}, \bibinfo{author}{{Hu}, W.}, \bibinfo{author}{{Liu}, Y.}, \bibinfo{year}{2013}.
\newblock \bibinfo{title}{{Projectile remnants in central peaks of lunar impact craters}}.
\newblock \bibinfo{journal}{Nature Geoscience} \bibinfo{volume}{6}, \bibinfo{pages}{435--437}.
\newblock \DOIprefix\doi{10.1038/ngeo1828}.

\end{thebibliography}

\end{document}